\documentclass[aps,prd,superscriptaddress,showpacs, twocolumn]{revtex4-1}
\usepackage{graphicx}
\usepackage{epstopdf}
\usepackage{amsmath}
\usepackage{amsfonts}
\usepackage{amssymb}
\usepackage{latexsym}
\usepackage{hyperref}
\usepackage[english]{babel}
\usepackage[utf8]{inputenc}
\usepackage[colorinlistoftodos]{todonotes}
\usepackage{xcolor}
\usepackage{slashed}
\usepackage{bm}  

\begin{document}
\title{Lab-based limits on the Carroll-Field-Jackiw Lorentz-violating electrodynamics}

\author{Y.M.P. Gomes}\email{ymuller@cbpf.br}
\affiliation{Centro Brasileiro de Pesquisas F\'{i}sicas (CBPF), Rua Dr Xavier Sigaud 150, Urca, Rio de Janeiro, Brazil, CEP 22290-180}

\author{P.C. Malta}\email{malta@thphys.uni-heidelberg.de}
\affiliation{Centro Brasileiro de Pesquisas F\'{i}sicas (CBPF), Rua Dr Xavier Sigaud 150, Urca, Rio de Janeiro, Brazil, CEP 22290-180}
\affiliation{Institut f\"ur theoretische Physik, Universit\"at Heidelberg, Philosophenweg 16, 69120 Heidelberg, Germany}

\begin{abstract}
The CPT-odd and Lorentz-violating Carroll-Field-Jackiw modification of electrodynamics is discussed and we study its effects on the energy spectrum of hydrogen, as well as in the generation of a momentum-dependent electric dipole moment for charged leptons. We also briefly comment on the possibility of the detection of Lorentz violation in measurements of vacuum birefringence in resonant cavities. The bounds found are based on local laboratory experimental limits and are not competitive with the ones coming from astrophysical considerations.
\end{abstract}

\maketitle


\section{Introduction} \label{sec_intro}

\indent

Despite its great success, the Standard Model (SM) of particle physics cannot be the final description of Nature and it has been shown that in some of its extensions, string theory, for example, it is possible that Lorentz symmetry is violated \cite{Kost,Roiban}. The observation of any, though feeble, signal of Lorentz symmetry violation (LSV) would represent a great breakthrough and demand the re-examination of the very basis of modern physics, i.e. theory of relativity and quantum field theory \cite{Liberati, Mattingly}.

A possible realization of LSV is achieved by considering a Lagrangian model where a field with non-trivial spin acquires a non-zero vacuum expectation value -- see e.g. ref.\cite{Kost}. In view of this work, one can introduce general non-dynamical tensors \cite{Kost2} and exploit several different couplings to the matter and gauge sectors of the SM \cite{Kost3}-\cite{Helayel1}. For a review of theory and experimental tests of CPT- and Lorentz-invariance, see refs.\cite{Mattingly, tables}. In the present paper we investigate the case of a LSV background 4-vector coupled directly to the photon sector, thus leaving the lepton sector untouched.

An interesting prospect to implement LSV in the (1 + 3) Maxwell sector was originally proposed by S. Carroll, G. Field and R. Jackiw \cite{CFJ} through the following CPT-odd Chern-Simons-like Lagrangian \cite{Kost2, Mewes_Kost}
\begin{equation}
\mathcal{L}_{\rm CFJ} = k_{\rm AF}^{\mu}A^{\nu}\tilde{F}_{\mu\nu}, \label{CFJ}
\end{equation}
where $A^{\mu}=\left(\phi, \, {\bf A} \right)$ is the usual 4-potential and $\tilde{F}^{\mu\nu}=\frac{1}{2!}\epsilon^{\mu\nu\alpha\beta}F_{\alpha\beta}$ is the dual of the electromagnetic field-strength tensor (we adopt $\epsilon^{0123} = +1$). In calculations we shall adopt $k_{\rm AF}^{\mu} \equiv k_{\rm AF} \, n^{\mu}$, where the coupling $k_{\rm AF}$ has canonical dimension of mass, while $n$ is dimensionless.

This term is gauge-invariant if $\partial_{\mu}n_{\nu} = 0$, that is, $n$ is a constant 4-vector, thus providing a preferred direction in space-time, i.e., a background, and breaking Lorentz invariance. It is possible, nevertheless, to give it a dynamic nature, where it may be interpreted as a pseudo-scalar field -- see e.g. \cite{Itin, CarrollField}. It has also been shown that this term can be radiatively generated when fermions couple to the electromagnetic field via a non-minimal covariant derivative, $D_{\mu} = \partial_{\mu} + ieA_{\mu} - \gamma_{5}b_{\mu}$. The constant 4-vectors $b^{\mu}$ and $n^{\mu}$ are therefore related, but the exact numerical relation between them has been long debated \cite{Klinkhamer}-\cite{Perez}.

Usual QED augmented by the CFJ Lagrangian is essentially a subset of the so-called minimal Standard Model Extension \cite{Liberati, Kost2}. Some of the classical features of this particular scenario were studied e.g. in ref.\cite{Adamm}, where it was shown that the CFJ interaction (also with a non-zero Proca mass term \cite{UFMA}) with a pure space-like background is stable, unitary and preserves causality, whereas time- and light-like backgrounds are potentially problematic \cite{CFJ, Klinkhamer, UFMA, Helayel3}. A space-like background is therefore the only healthy scenario available.

An important remark is in order at this point: the considerations above apply to a truly fixed, time-independent, background. These requirements are only explicit in an inertial reference frame, which is not the case of the Earth due to its sideral and orbital motions: in the lab the background would seem to rotate. A convenient and approximately inertial frame is, for example, the one attached to the Sun -- the so-called Sun-centered frame (SCF) \cite{Mewes} -- which is broadly used in the literature \cite{tables}.

In order to translate the accessible, but time-dependent, background as observed on Earth, $n_{\rm lab}$, in terms of combinations of the constant $n_{\rm Sun}$, we employ a general Lorentz transformation, i.e., $n_{\rm lab}^{\mu} = \Lambda^{\mu}_{\, \, \, \nu} \, n_{\rm Sun}^{\nu}$, where $\Lambda^{\mu}_{\, \, \, \nu}$ is given in ref.\cite{Mewes}. If we ignore sub-dominant boost effects of order $\beta \lesssim 10^{-4}$, we may write $n^{0}_{\rm lab} = n^{T}_{\rm Sun} \equiv 0$ and $n^{i}_{\rm lab} = R^{i \, J} (\chi, T_{\oplus})\,  n^{J}_{\rm Sun}$, where the rotation matrix is explicitly time-dependent ($T_{\oplus}$ is the time in the SCF).

Since experiments are usually conducted over long time-scales, the LSV signatures observed in Earth-bound experiments would be, thus, an effective time averages. The only non-vanishing (time-averaged) spatial components are then $n_{\rm lab}^{x} = -\sin\chi \, n_{\rm Sun}^{Z}$ and $n_{\rm lab}^{z} = \cos\chi \, n_{\rm Sun}^{Z}$, with $\chi$ the co-latitude of the experiment. As discussed below, the effects we consider in this paper are linear in $k_{\rm AF} \, n_{\rm lab}$, therefore only the $x$- and $z$-components of the background in the lab will be relevant to our analyses. Both may be expressed in terms of $k_{\rm AF} \,n_{\rm Sun}^{Z}$ and our goal is precisely to constrain it.

The CFJ Lagrangian above would induce optical effects during the propagation of radiation through vacuum (see Section \ref{sec_cav}), and Carroll, Field and Jackiw used data on the rotation of the plane of polarization from distant galaxies in order to impose strong limits on $k_{\rm AF}$. Given that no significant evidence of such effects was found, they could set a tight upper bound on the LSV parameter, namely, $k_{\rm AF} < 10^{-42} \, {\rm GeV}$ \cite{CFJ, CarrollField,Goldhaber}. Limits on this parameter have been searched for in many contexts, mainly astrophysical, e.g. CMB polarization \cite{Kost5, Kost6}, and are currently as strict as $k_{\rm AF}  \lesssim 10^{-43} \, {\rm GeV}$ (see ref.\cite{tables}, Table D12, and references therein).

Here we apply eq.\eqref{CFJ} to systems available at much shorter distance scales, where Earth-bound laboratory experiments may be used to constrain the predicted LSV effects. This is a valid effort, given that the apparatus is under the experimeter's control, as opposed to cosmological or astrophysical tests where sizeable uncertainties may arise due to complicated models describing the interstellar medium and light propagation therein. We discuss LSV effects in the context of the CFJ modification of QED in two main fronts: energy shifts in the spectrum of the hydrogen atom and the generation of an electric dipole moment for charged leptons. We also briefly address measurements of rotation in the polarization of light in resonant cavities.

As we shall see, the LSV-induced corrections to the Coulomb potential appear already at tree-level via velocity- and spin-dependent interaction potentials. For the leptonic electric dipole moment ($\ell$EDM), on the other hand, it is necessary to compute the one-loop correction to the corresponding form factor, which is found to be explicitly momentum-dependent, thus allowing for an enhancement in high energies. Despite of this interesting feature, we expect it to remain unaccessable to experiment in the foreseable future. Resonant cavities would present, nonetheless, a good prospect to perform local tests on LSV and to potentially constrain $k_{\rm AF} \,n_{\rm Sun}^{Z}$ even further.

This paper is organised as follows: in Section \ref{sec_pots} we discuss the interparticle potential between leptons and apply it to the hydrogen atom and, in Section \ref{sec_edm}, we calculate the one-loop CFJ contribution to the $\ell$EDM. In Section \ref{sec_cav} we address some classical features of the model and connect it to a resonant cavity. Section \ref{sec_conc} is devoted to our concluding remarks. We use natural units ($c = \hbar = 1$) throughout.


\section{Interparticle potential} \label{sec_pots}

\indent

The CFJ parity-odd Lagrangian, eq.\eqref{CFJ}, modifies the quadratic piece of the usual Maxwell sector in QED, therefore altering the propagator of the photon. This modification entails that photon-mediated interactions will necessarily include a (small) LSV signature, possibly giving rise to anisotropies involving the fixed background.

The relatively high precision attained in spectroscopy experiments motivates us to consider the effect of the CFJ corrections to the Coulomb potential in the study of atomic systems, the simplest of which is the hydrogen atom. To do so, one needs to compute the potential (operator) between the proton -- here treated as a point-like, structureless, fermion -- and the electron. The interaction of two spin-1/2 fermions can be treated in the non-relativistic (NR) regime through the concept of interparticle potential, which is given by the first Born approximation \cite{Dobrescu}, $V(r) = - \int\frac{d^{3}{\bf q}}{(2\pi)^{3}}\mathcal{M}({\bf q})e^{i{\bf q} \cdot {\bf r}}$, where $\mathcal{M}$ is the NR amplitude, ${\bf q}$ is the mometum transfer and ${\bf r}$ is the relative position vector.

The one-photon exchange amplitude may be schematically written as $\mathcal{M} \sim J_1^{\mu} \langle A_{\mu}A_{\nu} \rangle J_2^{\nu}$, where $J_{1,2}^{\mu}$ represent the contraction of the on-shell (external) spinors and the $\ell\ell\gamma$ vertex, and $\langle A_{\mu}A_{\nu} \rangle$ is the effective photon propagator. Given that LSV effects are expected to be tiny, we do not use the full propagator \cite{UFMA}, but merely consider the CFJ term \eqref{CFJ} as a true bilinear interaction in the photon sector, i.e., an effective vertex to be inserted into the usual QED propagator, $\langle A_{\mu}A_{\nu} \rangle = \frac{-i\eta_{\mu\nu}}{p^{2} + i\varepsilon}$. Under these assumptions one may write the CFJ vertex as \cite{Daniel}
\begin{equation}
V^{\mu\nu}_{\gamma\gamma} = 2 \left( k_{\rm AF} \right)_{\alpha} \epsilon^{\mu\alpha\beta\nu} p_{\beta}, \label{vi}
\end{equation}
while the QED tree-level vertex remains unaltered and reads $V^{\mu}_{\ell\ell\gamma} = ie\gamma^{\mu}$.

We consider the interaction in the center of mass frame, in which fermion ``1" has incoming and outgoing momenta denoted by $p = P + q/2$ and $p' = P - q/2$, respectively, where $P$ is the average mometum and $q$ is the momentum carried by the virtual photon. Similar definitions hold for fermion ``2" (with $p \rightarrow - p$ and $p' \rightarrow - p'$). Applying the Feynman rules and noting that $q_{\mu}\left[ \bar{u}\gamma^{\mu}u \right]_{1,2}=0$ for the conserved external currents, we obtain
\begin{equation}
i\mathcal{M} = \frac{2e_{1}e_{2} k_{\rm AF}}{(q^{2})^{2}} \left[ \bar{u}\gamma^{\sigma}u \right]_{1} \epsilon_{\sigma\alpha\beta\rho}n^{\alpha}q^{\beta}    \left[ \bar{u}\gamma^{\rho}u  \right]_{2}, \label{amps}
\end{equation}
where we shall assume an elastic interaction, $q^{\mu}=(0,{\bf q})$. As discussed in Section \ref{sec_intro}, the background above is the one measured in the lab, i.e., ${\bf n} = {\bf n}_{\rm lab}$, and, for the sake of simplicity, we shall only transform to the SCF variables in the end of our calculation.


In the NR limit the current for fermion 1 has components $\left[\bar{u}\gamma^{0}u\right]_{1} \sim 1$ and $\left[\bar{u}\gamma^{i}u\right]_{1} = \frac{{\bf P}_{i}}{m_{1}} - \frac{i}{2m_{1}}\epsilon_{ijk} \, {\bf q}_{j}\langle \sigma_{k} \rangle_{1}$, with similar results for current 2, provided one makes the appropriate changes in momenta (${\bf P} \rightarrow - {\bf P}$ and ${\bf q} \rightarrow - {\bf q}$). In our notation $\langle \sigma \rangle_{1,2}=\chi^{\dagger}\sigma_{1,2}\chi$, with $\chi$ being the basic spin-up or -down spinor satisfying $\chi_{a}^{\dagger}\chi_{b}=\delta_{ab}$, and $\sigma_{1,2}$ the usual Pauli spin matrices acting on particles 1 and 2, respectively. 


We now plug eq.\eqref{amps} into the definition of $V(r)$ in order to obtain the following potentials,
\begin{eqnarray}
V_{{\bf P}}(r) & = & \frac{\alpha \, k_{\rm AF}}{\mu r} \left( {\bf n} \cdot {\bf L} \right) \label{pot1} \\
V_{\sigma}(r) & = &  \frac{\alpha \, k_{\rm AF}}{2m_{1}m_{2}r} \left[ m_{1}{\bf n}\cdot\langle \sigma \rangle_{2}  +
m_1 ({\bf n}\cdot \hat{{\bf r}})\left( \hat{{\bf r}}\cdot\langle \sigma \rangle_{2} \right) \right] + \nonumber \\
& + & 1 \leftrightarrow 2,     \label{pot2}
\end{eqnarray}
where $\mu$ is the reduced mass of the system and ${\bf L}={\bf r} \times {\bf P}$ is the orbital angular momentum. The electric charges were set as $e_{1} = -e_{2} = \sqrt{4\pi \alpha}$, with $\alpha$ being the electromagnetic fine structure constant. The final result is just the sum of eqs.\eqref{pot1} and \eqref{pot2}: $\delta V_{C}(r)= V_{{\bf p}}(r) + V_{\sigma}(r)$, and it represents an additional LSV contribution to the well-known Coulomb interaction between two charges. We note, furthermore, the pseudo-scalar character of these potentials -- a clear sign of their CPT-odd origins.


The potentials above are spin- and velocity-dependent and could produce interesting consequences at the macroscopic level, inducing deviations in the dominant Coulomb force in the form of possible angle-dependent corrections to the inverse-square law observable in experiments involving e.g. spin-polarized objects \cite{Adelberger}. If, for simplicity, we consider instead the interaction energy between two charged but unpolarized bodies as given by $\delta V_C (r) = V_{{\bf P}}(r)$, we may extract the force per interacting pair of particles as $- {\bm \nabla} \delta V_C (r)$, that is 
\begin{equation}
{\bf f}_{\rm LSV}  =  \frac{\alpha k_{\rm AF}}{m_{1}r} \left[ \left( {\bf n}\times{\bf P} \right) + \frac{1}{r}\left( {\bf n}\cdot{\bf L} \right)\hat{{\bf r}}  \right],     \label{force}
\end{equation} 
where we assumed that $r$ is much greater than the typical dimensions of bodies ``1" and ``2" and that body ``2" is stationary ($m_2 \gg m_1$) and centered at the origin. The total force would be $\mathcal{N}_{\rm eff}{\bf f}_{\rm LSV}$, where $\mathcal{N}_{\rm eff}$ describes the effective number of interacting particles. This force would act as a small velocity-dependent perturbation to the dominating Coulomb (and gravitational) interaction between the two electrically charged objects.

The first term in eq.\eqref{force} represents a precession of the 3-momentum ${\bf P}$ around the axis defined by the fixed background ${\bf n}$. To see this it suffices to consider that ${\bf P}\cdot \frac{d{\bf P}}{dt} = 0$, so that the module of ${\bf P}$ is constant, i.e., time-independent. Similarly, the angle given by $\cos \vartheta = \frac{{\bf n}\cdot{\bf P}}{|{\bf n}||{\bf P}|}$ is also fixed in time (for small periods where the time-dependence of ${\bf n}$ itself can be neglected), so that the 3-momentum circles around the direction of ${\bf n}$. The second term shares more similarities with the typical Coulomb force, since it is radial and decays with the inverse square of the distance, but contains an unusual dependece on angular momentum, which also controls whether this term is attractive or repulsive. Besides the discussion above in terms of the LSV-originated force on charged leptons, the interaction from eq.\eqref{POT} may also induce a spontaneous torque on a pair of charges \cite{Helayel2}.

We now turn to our main interest: the application of our results, eqs.\eqref{pot1} and \eqref{pot2}, to the hydrogen atom. Given that the proton is a thousand times heavier than the electron, $\delta V_{C}(r)$ reads
\begin{equation}
\delta V^{\rm H}_{\rm C}(r) = \frac{\alpha k_{\rm AF}}{m_{e}r} \left[ {\bf n}\cdot{\bf L} + \frac{1}{2}{\bf n}\cdot{\bf \langle \sigma \rangle} + \frac{1}{2}\left( \hat{{\bf r}}\cdot{\bf n}  \right)\left(\hat{{\bf r}}\cdot\langle \sigma \rangle \right)  \right]  \label{POT}
\end{equation}
which represents a Lorentz violating CPT-odd correction to the electron-proton electromagnetic interaction. According to usual quantum-mechanical time-independent perturbation theory, in order to evaluate the first-order energy shift associated with this perturbation, we need to calculate $\Delta E_{\rm LSV}^{\rm H} = \langle \psi^0 | \delta V^{\rm H}_{\rm C} | \psi^0 \rangle$, with $|\psi^0\rangle$ being adequate eigenstates of the free hydrogen atom.

Since the problem involves not only the orbital angular momentum, but also the spin degrees of freedom, we need to build the angular wave functions for the total angular momentum, ${\bf J} = {\bf L} + {\bf S}$, which are given below for the case of a generic orbital angular mometum ${\bf L}$ coupled to a spin-1/2:
\begin{eqnarray}
\Theta_{j=\ell + \frac{1}{2}}(\theta,\phi) & = & \sqrt{\frac{\ell + m_{\ell} + 1}{2\ell + 1}}Y_{\ell, \, m_{\ell}}(\theta,\phi)\,\chi_{_{+}} + \nonumber \\
& + & \sqrt{\frac{\ell - m_{\ell}}{2\ell + 1}}Y_{\ell, \, m_{\ell}+1}(\theta,\phi)\,\chi_{_{-}}  \\
\Theta_{j=\ell - \frac{1}{2}}(\theta,\phi) & = & \sqrt{\frac{\ell - m_{\ell}}{2\ell + 1}}Y_{\ell, \, m_{\ell}}(\theta,\phi)\,\chi_{_{+}} + \nonumber \\
& - & \sqrt{\frac{\ell + m_{\ell} + 1}{2\ell + 1}}Y_{\ell, \, m_{\ell}+1}(\theta,\phi)\,\chi_{_{-}} 
\end{eqnarray}
both with $m_j = m_{\ell} + 1/2$. The final normalized wave functions are then $\psi^0(r,\theta,\phi) = R_{n,\ell}(r)\Theta_j (\theta,\phi)$, where $Y_{\ell,m_{\ell}} (\theta,\phi)$ and $R_{n,\ell}(r)$ are the well-known spherical harmonics and radial function for the hydrogen atom, and $\chi_{_{\pm}}$ are the spin eigenfunctions. Here, $n$ -- not to be confused with the background, $\ell$ and $m_{\ell}$ are the principal, angular and azimuthal quantum numbers, respectively.

As discussed in Section \ref{sec_intro}, after averaging, the background as seen in the lab is given by ${\bf n} = {\bf n}_{\rm lab} = (n^{x},0,n^{z})$, where we omit the sub-script for convenience. With this, the total energy shift is given by $\Delta E_{\rm LSV}^{\rm H} = \Delta E_1 + \Delta E_2 + \Delta E_3$, where
\begin{eqnarray}
\Delta E_1 & = & \mathcal{G} \left[ n^{x}\langle L_x + \frac{1}{2}\sigma_x \rangle + n^{z}\langle L_z + \frac{1}{2}\sigma_z \rangle  \right] \label{DE1} \\
\Delta E_2 & = & \frac{\mathcal{G} n^{x}}{2}  \langle \sin\theta \cos\phi  \left(\hat{{\bf r}}\cdot \sigma \right) \rangle  \label{DE2} \\
\Delta E_3 & = & \frac{\mathcal{G} n^{z}}{2} \langle \cos\theta  \left(\hat{{\bf r}}\cdot \sigma \right) \rangle  \label{DE3}
\end{eqnarray}
with $\mathcal{G} = \frac{\alpha k_{\rm AF}}{m_{e}} \overline{\left(r^{-1}\right)} = \frac{\alpha k_{\rm AF}}{m_{e}a_0 n^2}$ ($a_0 = 2.68 \times 10^{-4} \, {\rm eV}^{-1}$ is the Bohr radius). It is easy to check that, for $j = \ell + 1/2$, we have $\langle \psi^0 | L_z | \psi^0 \rangle = m_{\ell} + \frac{\ell - m_{\ell}}{2\ell + 1}$ and $\langle \psi^0 | \sigma_z /2 | \psi^0 \rangle = \frac{m_{\ell} + 1/2}{2\ell + 1}$, so that 
\begin{equation}
\Delta E_1 = \frac{\alpha k_{\rm AF} \, n^{z}_{\rm lab}}{m_{e}a_0 n^2} \left(m_{\ell} + 1/2 \right), \label{e_1}
\end{equation}
where we used that the contribution proportional to $n^{x}$ is automatically zero due to the orthogonality of the functions involved. Similar arguments lead to $\Delta E_2 = 0$.


Finally, $\Delta E_3$ may be written as $\Delta E_3 = \frac{\alpha k_{\rm AF} \, n^{z}_{\rm lab}}{2m_e a_0 n^2} \delta\mathcal{E}_3$, cf. eq.\eqref{DE3}, and, after employing the algebra of angular momentum \cite{Arfken}, we find $\delta\mathcal{E}_3 = \frac{2(m_{\ell} + 1/2)}{(2\ell + 1)(2\ell + 3)}$, so that our final result reads
\begin{equation}
\Delta E_{\rm LSV}^{\rm H} = \frac{4\alpha k_{\rm AF} \, n^{z}_{\rm lab}}{m_{e}a_0} \frac{m_{\ell} + 1/2}{n^2} \frac{\left( \ell + 1 \right)^2}{(2\ell + 1)(2\ell + 3)}, \label{espec}
\end{equation}
with a similar expression for the $j = \ell - 1/2$ case.

The quantity obtained above represents the energy shift to the spectral lines of hydrogen due to Lorentz violating effects. The aforementioned spectrum is known to a high level of accuracy and the fact that no deviations have been found allows us to place an upper bound on the magnitude of $\Delta E_{\rm LSV}^{\rm H}$. Optimistically, we may use the currently best precision in spectroscopic measurements, $\epsilon_{\rm exp}^{\Delta E^{\rm H}} = 4.2 \times 10^{-15} \, {\rm eV}$ \cite{Parthey}, and demand that $\Delta E_{\rm LSV}^{\rm H} < \epsilon_{\rm exp}^{\Delta E^{\rm H}}$, i.e., we demand that the LSV effect lies below experimental uncertainty. From this requirement we obtain the upper bound ($k_{\rm AF}^{\mu} \equiv k_{\rm AF} \, n_{\rm Sun}^{\mu}$)
\begin{equation}
k_{\rm AF}^{Z} \lesssim  10^{-19} \, {\rm GeV} \label{limit_1}
\end{equation}
at the $1 \, \sigma$ level \cite{Parthey}


\section{Electric dipole moment} \label{sec_edm}

\indent 

If an elementary particle possesses a non-zero electric dipole moment, ${\bf d}$, it has to point in the direction of its spin, since this is the only vector available in the rest frame of the particle. When placed in an external electric field the particle will be subject to an interaction of the form $-{\bf d}\cdot{\bf E}$, which can be recast as $-d ({\bf S}\cdot{\bf E})$, and this interaction term violates both P- and T-symmetries. Standard QED, on the other hand, is parity-invariant, so that such an electric dipole interaction cannot be described by pure QED processes, that is, $d^{QED} \equiv 0$.

Within the Standard Model it is possible to generate a small leptonic EDM when strong and electroweak interactions are taken into account \cite{Pospelov}-\cite{Susuki}. For the electron, its theoretical magnitude is bounded by $|d_{e}^{\rm SM}| < 10^{-38} \, {\rm e\cdot cm}$, while the best experimental limit is $|d^{\rm exp}_{e}| < 8.7 \times 10^{-29} \, {\rm e\cdot cm}$, at $90\%$ CL \cite{ACME2}. The relatively strong experimental bounds on $d_{e}$ can be used as a means to extract limits on the physical properties of new particles, such as axions \cite{Mantry, Hill}, supersymmetric particles \cite{Nath}, Majorana neutrinos \cite{Krause2} and dark matter \cite{Nelson}.

We shall now turn to the actual calculation of the LSV contribution to $d_{\ell}$. It is clear that the tree-level contribution to the $\ell$EDM is zero in the CFJ scenario -- the tree-level QED $\ell\ell\gamma$ vertex remains unaltered -- so we must look at higher orders.

\begin{figure}[htb!]
\centering
\includegraphics[height=5.7cm, angle = 0]{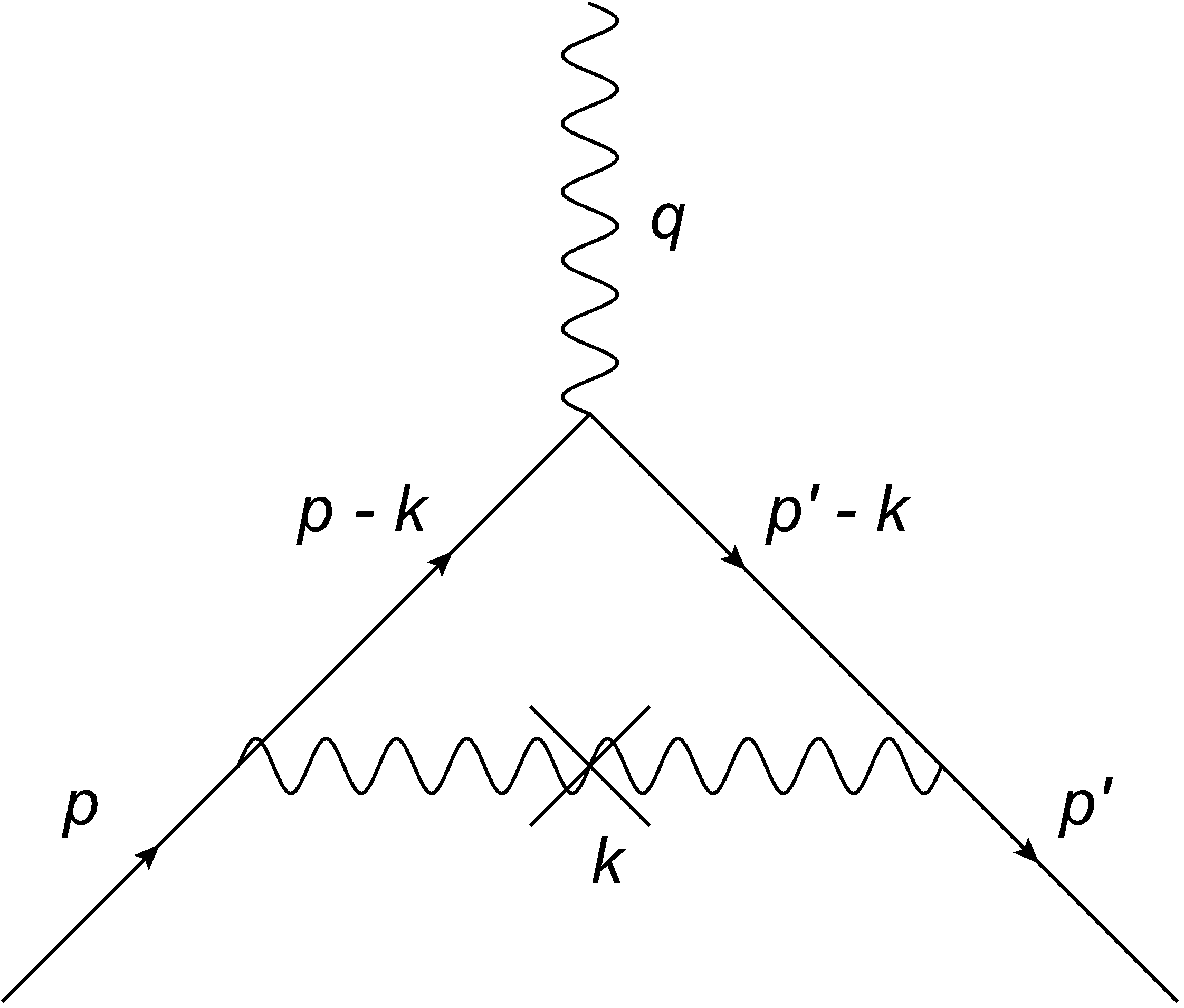}
\caption{\label{fig:overleaf}Vertex structure and momenta attributions; the cross indicates the vertex insertion.} \label{Fig1}
\end{figure}

The first non-zero contribution comes from the one-loop vertex correction diagram, as shown in fig.\eqref{Fig1}. Following the momentum assignments we have 
\begin{equation}
\Lambda_{\mu}(p,p',q) =  -2e^{2}k_{\rm AF}\epsilon^{\nu\alpha\beta\rho}n_{\alpha} \times I_{\beta\nu\mu\rho}(p,p',q),
\end{equation}
where
\begin{equation}
I_{\beta\nu\mu\rho} = \int \frac{d^{4}k}{(2\pi)^{4}} \frac{\gamma_{\nu} \left( \slashed{p'} - \slashed{k} + m_{\ell} \right) \gamma_{\mu} \left( \slashed{p} - \slashed{k} + m_{\ell} \right)\gamma_{\rho} k_{\beta}}{\left( k^{2} \right)^{2} \left[ (p' - k)^{2} - m_{\ell}^{2} \right] \left[ (p - k)^{2} - m_{\ell}^{2}  \right] },   \label{EDM1}
\end{equation}
and we observe that the superficial degree of divergence of this diagram is $-1$, meaning that it behaves as $\sim 1/k$ in the UV limit. Remembering that the corresponding diagram in usual QED, which describes the g-factor, displays a superficial logarithmic divergence, we conclude that the role of the vertex insertion is to reduce the degree of divergence and render the diagram UV-finite.


Since the integral in eq.\eqref{EDM1} is finite in $D=4$ there is no need to regularize it and we directly evaluate the vertex correction as $\Lambda_{\mu}(p,p',q) = -\frac{ie^{2}k_{\rm AF}}{64\pi^{2}m_{\ell}^{2}} \epsilon^{\nu\alpha\beta\rho}n_{\alpha} T_{\beta\nu\mu\rho}$, with the momentum-dependent object $T_{\beta\nu\mu\rho}$ being a complicated function involving products of up to five gamma matrices. We shall not give its expressions here. The vertex $\Lambda_{\mu}(p,p',q)$ is the LSV contribution to $d_{\ell}$ we were looking for, but, in order to extract it, we need to obtain the corresponding form factor.


The electromagnetic current can be decomposed as $\langle p'| J^{\mu}_{\rm em} | p \rangle = \frac{\sigma^{\mu\nu}\gamma_{5}}{2m_{\ell}}q_{\nu}F_{\rm edm}(q^2)  + \cdots$, where $F_{\rm edm}(q^2)$ is the desired form factor and the ellipsis denotes the other Lorentz structures and their respective form factors, which are not of interest here \cite{Paschos}. Above $q = p - p'$ is the momentum transfer. In this paper we consider only the LSV effects in the photon sector, so no other form factor other than $F_{\rm edm}(q^2)$ is relevant, as the (free) Dirac equation remains unaltered. The vertex function $\Lambda_{\mu}(p,p',q)$ plays the role of a LSV correction to the usual electromagnetic current, so that our task is to extract $F_{\rm edm}(q^2)$ and read the $\ell$EDM, which is then given by $d_{\ell} = -F_{\rm edm}(q^2 = 0)/2m_{\ell}$.

Obtaining $F_{\rm edm}(q^2)$ is cumbersome due to the complicated form of $\Lambda_{\mu}(p,p',q)$. It is possible, however, to simplify matters by applying an appropriate projector \cite{proj}: $\mathcal{P}^{\mu}_{\rm edm} =  \frac{im_{\ell}  \left( p + p' \right)^{\mu}}{q^4 - 4m_{\ell}^{2}q^2} \left[ \left( \slashed{p} + m_{\ell} \right) \gamma_{5} \left( \slashed{p'} + m_{\ell} \right)\right]$, which automatically selects the form factor we want. The projector above acts on the vertex correction and we obtain $F_{\rm edm}(q^2) = \text{Tr}\left[ \Lambda_{\mu}\mathcal{P}^{\mu}_{edm} \right]$, with the trace evaluated for external (on-shell) leptons, that is, $p^2=p'^2=m_{\ell}^{2}$ and $p\cdot p' = m^{2}_{\ell} - q^{2}/2$. At this point it is convenient to leave $q^2 \neq 0$ in order to extract the finite contributions from the trace above in the limit of massless photons.

This task may be performed in an automated fashion through the use of Hiren Patel's Package-X \cite{PX}, and the form factor is found to be:
\begin{equation}
F_{\rm edm} (q^2 = 0) = -\frac{e^2 k_{\rm AF}}{12\pi^2 m_{\ell}^2} \left[ p\cdot n - p'\cdot n \right] + \text{IR}, \label{FEDM}
\end{equation}
where $\text{IR}$ indicates infra-red terms. Such divergences appear as $1/x$ factors in the Feynman integrals ($x \rightarrow 0$) due to $m_{\gamma} = 0$ and as $1/q^2$ factors in the traces. The appearance of the latter may be interpreted as follows. We are considering the CFJ correction as a true vertex and not using the associated full (complete) propagator -- this is essentially equivalent to taking only the lowest-order term in $k_{\rm AF}/|{\bf q}|$ in the expansion of the complete propagator. However, the loop integration does not contemplate only high momenta, but also regions where $k_{\rm AF}/|{\bf q}| \ll 1$ may not be fulfiled, so we expect these divergences to vanish upon using the complete LSV-modified propagator.


Finally, by using that $q = p - p'$ and the definition of the EDM in terms of the associated form factor, we obtain:
\begin{equation}
d_{\ell} = \frac{\alpha k_{\rm AF}}{6\pi m_{\ell}^{3}} \left( q \cdot n \right), \label{EDMfinal}
\end{equation}
which shows that the $\ell$EDM is momentum-dependent. A simular effect was found in ref.\cite{Haghighat} for a different sector of the Standard Model Extension \cite{Kost2}, but with a quadratic momentum dependence.

It is interesting to note, though not surprising, that, in a space-time with a fixed non-dynamic background, the spin is not the only vector available to support the electric (or magnetic) moment of an elementary particle. Furthermore, in order to build the scalar $d_{\ell}$ we need another vector, and the only possibilities are $p$ and $p'$ -- in our case, the special combination given by $q =p - p'$. This may be interpreted in terms of an interplay between the background and the applied electromagnetic field, which carries the momentum transfer $q$, so that, together, they produce a non-vanishing $\ell$EDM, i.e., induce an asymmetry in the charge distribution of the lepton.

We note, however, that the form of the $\ell$EDM as given by eq.\eqref{EDMfinal} is not helpful from the experimental point of view: for an elastic interaction ($q^0 \approx 0$) with $|{\bf q}|^2 \ll m^2_{\ell}$ we have $d_{\ell} \sim 0$. Besides this, two aspects are specially relevant here: the nature of the measurements (e.g. as performed by the ACME collaboration \cite{ACME2}) and its time-scale. Let us first decompose the (say) initial momentum of the electron as ${\bf p} = {\bf p}_{m} + {\bf p}_{s}$, where ${\bf p}_{m,s}$ denote the components of the momentum relative to the molecule and the SCF, respectively. The first aspect is connected with the form of $d_{\ell} \sim {\bf p} \cdot {\bf n}$ and the fact that ACME's measurements were performed with (ThO) molecules, around which the electrons quickly revolve. Being bound to it, their momenta are also limited and, over time, average to zero, i.e., $\langle {\bf p}_m \rangle = 0$. Similar arguments would apply for ``free" relativistic leptons in storage rings \cite{Farley2, UAL}. This brings us to the second point.

Since ${\bf p}_m$ does not contribute, one should consider the general motion of Earth and the experiment attached to it relative to the SCF, cf. Section \ref{sec_intro}. The data from ACME's latest result was taken during $\sim 10$ days, but these were spread over months, and their analysis was not sensitive to such possible long-term time modulations. The momentum of the lab relative to the SCF is ${\bf p}_s \sim {\bm \beta}$, with the boost factor ${\bm \beta}$ given by eq.(C2) in ref.\cite{Mewes}, where it becomes clear that all components of ${\bf p}_s$ are periodic functions of time. Therefore, the time-averaged LSV effects $\sim \langle {\bm \beta} \rangle$ also vanish and the application of the upper limit on the eEDM \cite{ACME2} as a means to constrain the space-like LSV parameters is not possible.


%

In any case, in a speculative note, if we could use the bound on the LSV parameter given in ref.\cite{tables}, $k_{\rm AF} \sim  10^{-43} \, {\rm GeV}$, the energy (or momentum) necessary to reach the upper limit of $|d^{\rm exp}_{e}|$ would be $\sim 10^{21} \, {\rm GeV}$. This indicates that the CFJ contribution to the eEDM would only be sensible for extreme energies, around the Planck scale, $E_{\rm Planck} \simeq 10^{19} \, {\rm GeV}$, therefore remaining out of experimental reach for the foreseeable future (QED is also expected to break down at such high energy scale). This suggests that the CFJ model induces only very small effects and is, therefore, not responsible for a finite eEDM, should one be eventually found.

\section{Resonant cavities} \label{sec_cav}

\indent

The CFJ Lagrangian, eq.\eqref{CFJ}, is CPT-odd: in the case of a pure time-like background vector, $n^{\mu}=(n^{0}, 0)$, we obtain $\mathcal{L}_{\rm CFJ} \sim n^{0} \, {\bf B}\cdot{\bf A}$, while for the purely space-like case, $n^{\mu}=(0, {\bf n})$, a similar calculation shows that $\mathcal{L}_{\rm CFJ} \sim \phi \, {\bf n}\cdot{\bf B} - {\bf n}\cdot ( {\bf A} \times {\bf E} )$, with ${\bf E}$ and ${\bf B}$ the electric and magnetic fields, respectively. In the light of parity and time reflection transformations these terms show a clear pseudo-scalar character.

Adding eq.\eqref{CFJ} to the usual Maxwell kinetic term, $-\frac{1}{4}F_{\mu\nu}F^{\mu\nu}$, and varying the action with respect to $A^{\mu}$ gives us the LSV-modified Maxwell equations: $\partial_{\mu}F^{\mu\nu} = -2\left( k_{\rm AF} \right)_{\alpha} \tilde{F}^{\alpha\nu}$ (in the absence of matter sources). In momentum space, Gauss and Amp\`ere's laws become 
\begin{eqnarray}
{\bf k}\cdot{\bf E} & = & 2i{\bf n}\cdot{\bf B}  \label{max_1} \\
{\bf k}\times{\bf B} + \omega{\bf E} & = & -2i{\bf n}\times{\bf E}, \label{max_2}
\end{eqnarray}
where we temporarily absorbed $k_{\rm AF}$ in ${\bf n}$; $\omega$ and ${\bf k}$ are the 3-momentum and energy of the photon, respectively. The magnetic Gauss and Farady's induction laws remain untouched, since they stem from $\partial_{\mu}\tilde{F}^{\mu\nu} = 0$ \cite{CFJ}.

It is simple to see that ${\bf B}$ is simultaneously perpendicular to both ${\bf k}$ and ${\bf E}$, but, from the second equation above, one finds that the electric field is not purely transverse, but rather satisfies $( {\bf k} - \frac{2i}{\omega}{\bf n}\times{\bf k} )\cdot{\bf E} = 0$. This implies that the Poynting vector $\sim {\bf E}\times{\bf B}$ is not entirely parallel to the wave vector ${\bf k}$, a situation also encountered in e.g. electro-anisotropic uniaxial  media \cite{Wang}.

Another important information that may be obtained from the (general) Maxwell equations above concerns the dispersion relations. Working out the modified Maxwell equations one finds from the wave operators that the wave 4-vector satisfies
\begin{equation}
k^4 + 4k^2n^2 - 4(k\cdot n)^2 = 0, \label{disp}
\end{equation}
which, for the case of a space-like background, are approximately given by $\omega_{\pm} = |{\bf k}| \pm 2k_{\rm AF}\cos\psi + \mathcal{O}(k^2_{\rm AF})$, where $\cos\psi = {\bf k}\cdot{\bf n}/|{\bf k}||{\bf n}|$. This means not only that different modes propagate with different velocities, but also that the polarization plane is rotated by an amount $\Delta  \simeq k_{\rm AF} \, n_{\rm lab}^x L_x \cos\psi$ after traveling a length $L_x$ \cite{CFJ} (assuming that the experiment lies in the $xy$ plane in the reference frame of the lab).

A similar effect, Farady rotation, is observed whenever a linearly polarized wave passes through a dielectric exposed to an external magnetic field (aligned with the wave vector). Incidentally, if we express ${\bf E}$ and ${\bf B}$ in terms of the scalar and vector potentials, $\phi$ and ${\bf A}$, in the RHS of eqs.\eqref{max_1} and \eqref{max_2}, we arrive at LSV-modified Maxwell equations which are formally identical to its Lorentz-preserving counterpart in a dielectric medium, but here ${\bf P} \sim {\bf n} \times {\bf A}$ and ${\bf M} \sim \phi \, {\bf n}$ play the role of the polarization and magnetization, respectively \cite{Mewes}.

Experiments such as PVLAS \cite{PVLAS} use high finesse resonant cavities to search for the electromagnetic properties of the vacuum \cite{qed1, qed2} with intense lasers and are highly sensitive to rotations in polarization. Since at every reflection the direction of the propagation is inverted ($\cos\psi \rightarrow -\cos\psi$), the net rotation before and after reflection cancels on average; therefore, we use only one pass. Resonant cavities are usually designed to allow for the highest number possible of reflections (passes), therefore amplifying the effective optical path, but here we would like to obtain a rough estimate of the uncertainty in the measurement of polarization rotations over one single pass.

The final relative uncertainty on the rotation of polarization after $N \simeq 4.4 \times 10^5$ effective passes -- $N$ is the path-length amplification factor -- is $\epsilon_{\rm exp}^{\Delta} \sim 10^{-10}$ (at $1 \, \sigma$) \cite{PVLAS}, so that, for a single pass, we can estimate an uncertainty of $\epsilon_{\rm pass} \sim \sqrt{N} \times \epsilon_{\rm exp}^{\Delta} \simeq 6 \times 10^{-8}$. If we consider the PVLAS cavity with path length $L = 1.6 \, {\rm m}$ and a relative uncertainty of $\epsilon_{\rm pass}$ for the rotation in polarization, we may obtain an upper bound on $k_{\rm AF} \, n_{\rm lab}^x$ -- including a conservative factor of $\cos\psi \sim \mathcal{O}(10^{-1})$ -- by imposing that the LSV-induced birefringence in vacuum is smaller than $\epsilon_{\rm pass}$, i.e., $\Delta < \epsilon_{\rm pass}$.

With this simple assumption we may estimate that the LSV parameters in terms of the (time-averaged) SCF variable $k_{\rm AF}^Z \equiv k_{\rm AF} \, n_{\rm Sun}^Z$ (cf. Section \ref{sec_intro}) are bounded as 
\begin{equation}
k_{\rm AF}^Z \lesssim 8 \times 10^{-23} \, {\rm GeV}, \label{limit_2}
\end{equation}
whereby we note that more precise measurements and/or larger $L$ from e.g. the BMV experiment \cite{BMV} could potentially improve this upper limit and, hopefullly, eventually supersede the astrophysical bounds \cite{CFJ, tables}.

As discussed in Section \ref{sec_intro}, a time-like background brings theoretical difficulties -- this is the reason why we assumed $n^T_{\rm Sun} \equiv 0$ so far. However, if we insist on considering this possibility, we might be able to find stringent bounds on it. We work in the same approximation level as with the space-like components, i.e., we neglect effects of order $\beta \lesssim 10^{-4}$, so that $n^{0}_{\rm lab} = n^{T}_{\rm Sun} + \mathcal{O}(\beta)$ -- here the $\mathcal{O}(\beta)$ contributions are all time-dependent and are effectively washed away after time averaging \cite{Mewes}.

Working out the dispersion relation, eq.\eqref{disp}, for this specific case, we find that the two frequency modes induce a rotation in the polarization given by $\Delta \simeq k_{\rm AF} \, n_{\rm lab}^0 L$, where $L = ct$. This rotation does not depend on the projection of the linear momentum onto the (space-like) background, so there is no cancellation upon reflexion; we are then allowed to use $\epsilon_{\rm exp}^{\Delta} \sim 10^{-10}$. With this, we may estimate the following upper limit on a pure time-like LSV background ($k_{\rm AF}^T \equiv k_{\rm AF} \, n^T_{\rm Sun}$):
\begin{equation}
k_{\rm AF}^T \lesssim 10^{-25} \, {\rm GeV}, \label{limit_3}
\end{equation} 
which supports the theoretical indications that $k_{\rm AF}^T$ should be either exactly zero or extremely small \cite{CFJ, Klinkhamer, UFMA, Helayel3}.

Furthermore, we would like to note that this extrapolation could also be applied to the results in Section \ref{sec_pots}, but the limits on the time-component of the background would be essentially of the same order of magnitude as the one for the spatial components (cf. eq.\eqref{limit_1}). For this reason we refrain from re-doing the calculation explicitly for this case. Also, the conclusions in Section \ref{sec_edm} would not change by assuming a non-zero time-like component as we are in a regime where $q^0 \approx 0$, so we would still be unable to apply the experimental limits.


\section{Concluding remarks} \label{sec_conc}

\indent

In this paper we have studied a specific modification of standard QED, namely the Carroll-Field-Jackiw model given by eq.\eqref{CFJ}, in two different contexts: the interparticle potential between spin-1/2 fermions and the associated quantum-mechanical corrections to the spectrum of the hydrogen atom, the electric dipole moment of charged leptons at the one-loop level, as well as a brief application to resonant cavities, which incidentally provided the best upper bound on the LSV parameters. The bounds obtained are far less strict than that of ref.\cite{CFJ} and those listed in ref.\cite{tables}, but contrary to them, ours were extracted from local phenomena and experiments, therefore not depending upon astrophysical observations over cosmological distance scales and associated uncertainties.

Our study of the interparticle potential mediated by the LSV-modified propagator led us to spin- and velocity-dependent interactions which could interfere with the dominant Coulomb and gravitational forces between (un)polarized charged macroscopic objects. Next, we applied $\delta V_{\rm C}^{\rm H} (r)$ as a quantum-mechanical perturbation to the hydrogen atom, obtaining $\Delta E_{\rm LSV}^{\rm H}$, eq.\eqref{espec}, as a correction to the fine structure of the energy spectrum.

The background-dependent correction $\delta V_{\rm C}^{\rm H} (r)$, eq.\eqref{POT}, produces not only energy shifts in the spectrum, but may also induce changes in the (free) wave functions themselves. Such perturbed states ($| \psi^1 \rangle$) could give rise to other interesting effects, such as the generation of atomic electric dipole moments, $\langle \psi^1 | e {\bf R} | \psi^1 \rangle$, as well as induce non-zero quadrupole moments in the otherwise spherically symmetric ground state of the hydrogen atom \cite{Borges, Flambaum}. These are very interesting points and will be further examined and addressed elsewhere.

In Section \ref{sec_edm} we found that the background 4-vector may serve as support for a non-zero $\ell$EDM, which is also explicitly momentum-dependent, see eq.\eqref{EDMfinal}. However, due to the dependence of $d_{\ell}$ on $q = p-p'$, but not on the {\it average} momentum $P \sim p + p'$, and the experimental techniques used to measure it, we have not been able to set bounds on the LSV parameters.

Finally, our best bound on $k_{\rm AF}^{Z}$, eq.\eqref{limit_2}, was obtained from the non-observation of an LSV-induced vacuum birefringence, an effect analogous to Faraday rotation. We have not gone in the details of the cavity design, but rather outlined a general estimate. A closer analysis of the cavity operation and geometry would be able to refine it further, but our discussion indicates that this is a promising way to study not only non-linear properties of the vacuum predicted by QED or new beyond the Standard Model particles, e.g. axion-like particles and hidden photons \cite{JJ}, but also Lorentz violation and its induced effects on the electromagnetic vacuum \cite{Lipa} (for a more thourough overview see also ref.\cite{tables}, Table D12, and references therein).

\begin{acknowledgments}
The authors are grateful to J.A. Helay\"el-Neto, J. Jaeckel, H. Patel, N. Russel and V.A. Kosteleck\'y for interesting discussions and comments on the manuscript. This work was funded by the Brazilian National Council for Scientific and Technological Development (CNPq) and the German Service for Academic Exchange (DAAD). P.C.M. would also like to thank Elizabeth West and Brendon O’Leary for relevant comments regarding the results on the eEDM from the ACME collaboration.
\end{acknowledgments}




\end{document}